\documentclass[range]{ar2e}

\usepackage[dvips]{graphicx}
\usepackage{amssymb,amsfonts,amsmath,calc,ifsym,bbding,wasysym}
\usepackage[usenames]{color}
\usepackage{bm}
\usepackage[normalem]{ulem}
\usepackage{xr}
\usepackage{float}
\usepackage{chemarr,bbding,wasysym}
\usepackage{latexsym,amssymb,graphicx,amsmath,epsfig,calc,times}
\usepackage{chemarr,bbding,wasysym}
\newcommand{\Gcap}{G^\text{cap}}

\newcommand{\Fe}{F_\text{encap}}

\newcommand{\gb}{g_{\text{b}}}

\newcommand{\gsm}{g_\text{sm}}

\def\nnuc{{n_\text{nuc}}}
\def\tnuc{\tau_\text{nuc}}

\newcommand{\kt}{k_{\text{B}}T}
\def\fc{f_\text{c}}

\def\rhoTot{c_\text{T}}

\def\coreTot{x_\text{T}}

\def\rhoStar{c^*}
\def\rhoStarStar{c^{**}}

\def\chargeratio{r_\text{charge}}

\newcommand{\Gcore}{G^\text{core}}
\newcommand{\Ebend}{E_\text{bend}}

\newcommand{\Rh}{R_\text{H}}

\definecolor{Blue}{rgb}{0,0.0,1.0}
\definecolor{Red}{rgb}{1.0,0.0,0.0}
\definecolor{Green}{rgb}{0.0,0.35,0.0}
\definecolor{Grey}{rgb}{0.5,0.5,0.5}

\newcommand{\definition}[1]{}

\begin{document}



\jname{..}
\jyear{2000}
\jvol{}
\ARinfo{1056-8700/97/0610-00}

\title{Mechanisms of Virus Assembly}

\markboth{Perlmutter and Hagan}{Mechanisms of Virus Assembly}

\author{Jason D. Perlmutter, Michael F. Hagan
\affiliation{Martin Fisher School of Physics, Brandeis University, Waltham, MA, 02454; email: hagan@brandeis.edu}}


\begin{abstract}
Viruses are nanoscale entities containing a nucleic acid genome encased in a protein shell called a capsid, and in some cases surrounded by a lipid bilayer membrane. This review summarizes the physics that govern the processes by which capsids assembles within their host cells and in vitro. We describe the thermodynamics and kinetics for assembly of protein subunits into icosahedral capsid shells, and how these are modified in cases where the capsid assembles around a nucleic acid or on a lipid bilayer. We present experimental and theoretical techniques that have been used to characterize capsid assembly, and we highlight aspects of virus assembly which are likely to receive significant attention in the near future.\end{abstract}
\maketitle

\section{Introduction}
\label{sec:intro}
The formation of a virus is a remarkable feat of natural engineering. A large number ($\sim60-10,000$) of protein subunits and other components assemble from the crowded cellular milieu to form ordered, complete, reproducible structures on biologically relevant time scales. Viruses are infectious agents responsible for a significant portion of human diseases, as well as those of other animals, plants, and bacteria. Thus, it is of biomedical interest to understand their formation process, with the aim of designing antiviral therapies that block it, or alternatively reengineering viruses for use as targeted delivery vehicles. More generally, the assembly of basic units into structures with increased size and complexity is ubiquitous in biology and is playing increasingly important roles in nanoscience. Understanding the mechanisms by which viral components co-assemble may elucidate diverse classes of assembly reactions.

Viruses vary tremendously in size and complexity, ranging from the 16-nm satellite panicum mosaic virus (SPMV) \cite{Ban1995}, whose 826 nucleotide genome encodes for a single protein, to the $\mu$m-sized pandoravirus, whose 2.5 megabase genome is larger than some bacterial genomes and encodes for 2556 putative proteins \cite{Philippe2013}.  However, viruses share a common body plan, consisting of a genome, which can be single-stranded (ss) or double-stranded (ds) and can be RNA or DNA, surrounded by a protective container, which is usually a protein shell called a capsid. For `enveloped' viruses (e.g. HIV or influenza), the capsid is additionally surrounded by a lipid bilayer acquired from the host cell. Formation of an infectious virion requires assembly of the capsid, envelopment by a membrane (if enveloped), and packaging of the nucleic acid (NA) genome within. Many viruses with single-stranded genomes assemble spontaneously around their NA, as demonstrated in 1955 by the experiments of Fraenkel-Conrat and Williams in which tobacco mosaic virus RNA and capsid proteins spontaneously assembled into infectious virions in vitro \cite{Fraenkel-Conrat1955}. In contrast, the stiffness and high charge density of dsDNA or dsRNA preclude spontaneous encapsidation (unless the genome is first complexed with NA-folding proteins). Therefore, many dsDNA viruses assemble an empty protein capsid (procapsid) and a molecular motor that hydrolyzes ATP to pump the DNA into the capsid.

This review focuses on the physics of virus assembly --- how the interactions among proteins and other viral or non-viral components determine their assembly pathways and products. Although we focus on viruses, many aspects of these pathways and the factors that control them are generic to other biological or synthetic self-assembly reactions. We begin with a brief overview of virus structures, followed by a summary of the experimental and theoretical methods used to characterize the assembly process. Next, we describe the thermodynamics, kinetics, and underlying mechanisms associated with assembly of proteins into empty capsids (section~\ref{sec:empty}).  We then consider how this process is modified when the proteins assemble around NAs (section~\ref{sec:cargo}) or on lipid bilayer membranes (section~\ref{sec:membrane}).  We then briefly discuss how small molecules that modulate assembly mechanisms form a promising new class of antiviral agents (section~\ref{sec:antiviral}). In the latter three sections, we concentrate on recent results which were not covered in previous reviews on capsid assembly \cite{Mateu2013,Hagan2014}. Finally, we conclude by discussing open questions and possible future avenues of research.

\subsection{Capsid architectures}
\label{sec:anatomies}
\begin{figure} [hbt!]
\begin{center}
\epsfxsize=0.6\textwidth\epsfbox{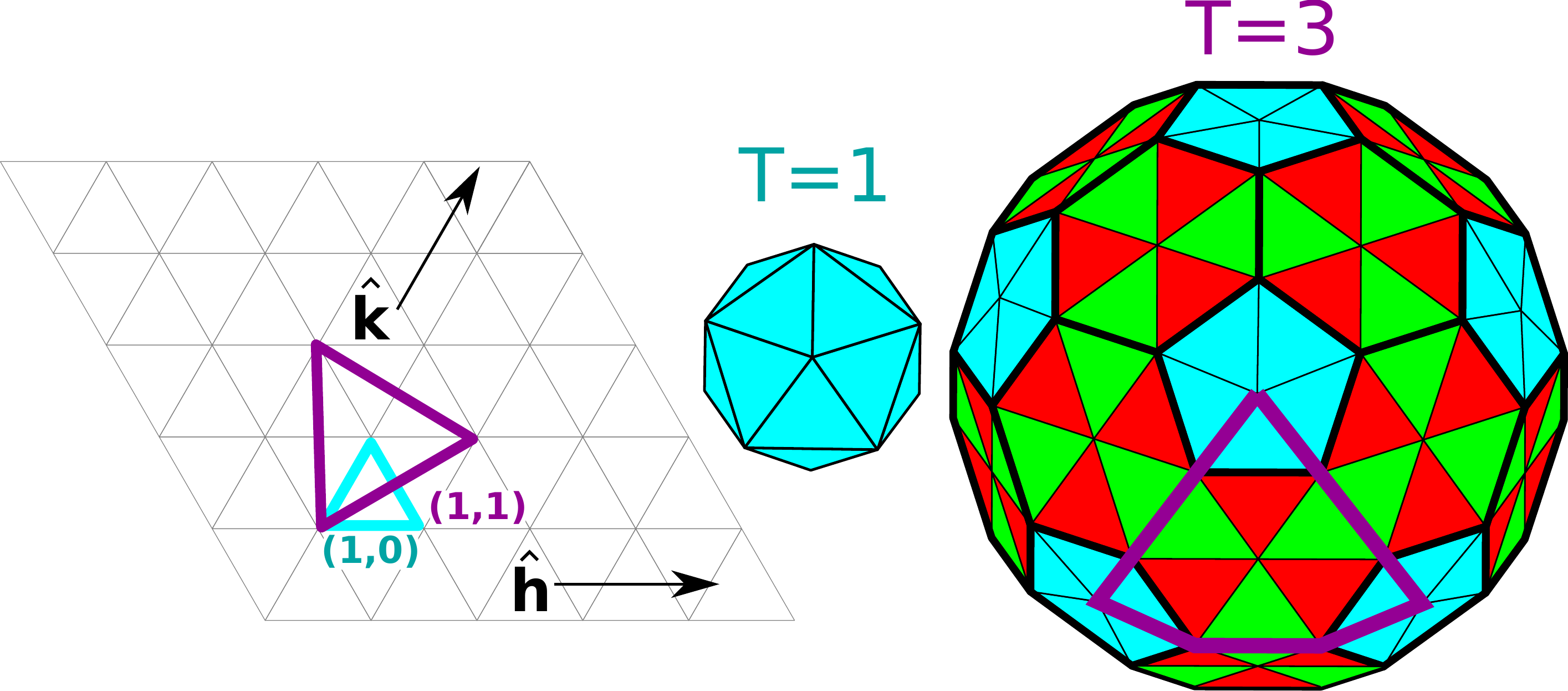}
\caption{  The geometry of icosahedral lattices. Moving $h$ and $k$ steps along each of the $\mathbf{\hat{h}}$ and $\mathbf{\hat{k}}$ lattice vectors results in a triangle with area $T/4$ (for unit spacing between lattice points), where $T$ is the triangulation number defined as $T=h^2+hk+k^2$.  The blue and purple triangles correspond to $T{=}1$ and $T{=}3$ respectively. The resulting icosahedrons are shown in the center and right images, with triangular facets in distinct (quasi-equivalent) environments distinguished by color. The purple triangle from the left image is inscribed on the $T{=}3$ icosahedron.
\label{fig:tNumber}
}
\end{center}
\end{figure}

The viral genome length, and hence the number of unique proteins that it can encode, is constrained by the requirement that it be enclosed by its capsid. Most capsids therefore comprise one or a few protein sequences arranged with a high degree of symmetry. The majority of viruses can be classified as rodlike or spherical, with the capsid proteins of rodlike viruses arranged with helical symmetry around the nucleic acid, and the capsids of most spherical viruses arranged with icosahedral symmetry. The number of units in a helix is arbitrary, and thus a helical capsid can accommodate a nucleic acid of any length. In contrast, icosahedral capsids are constrained by the fact that at most 60 identical subunits can form a regular polyhedron. Based on the observation that many capsids contain integer multiples of 60 proteins, Caspar and Klug  \cite{Caspar1962} proposed geometrical arguments that describe how multiples of 60 proteins can be arranged with icosahedral symmetry, where individual proteins interact through the same interfaces but take slightly different, or quasi-equivalent, conformations (reviewed in \cite{Johnson1997,Zlotnick2005}). A complete capsid is comprised of $60T$ subunits, where $T$ is the `triangulation number', which is equal to the number of distinct subunit conformations (Fig.~\ref{fig:tNumber}). An extensive collection of capsid structures determined from x-ray crystallography and/or cryo-electron microscopy (cryo-EM) data can be found at the VIPER website (http://viperdb.scripps.edu) \cite{Reddy2001}.

\subsection{Experimental and theoretical methods to characterize capsid assembly}
\label{sec:measurements}
{\bf Bulk experiments.} Capsid assembly kinetics  have been measured in vitro with size exclusion chromatography (SEC), small angle X-ray scattering (SAXS), and light scattering (e.g. \cite{Prevelige1993,Zlotnick1999,Zlotnick2000,Casini2004,Chen2008,Berthet-Colominas1987,Kler2012,Kler2013}, Fig.~\ref{fig:Nguyen}A below). The SEC experiments show that under optimal assembly conditions the only species present in detectable concentrations are either complete capsids or small protein oligomers which we refer to as the basic assembly unit. The size of the basic assembly unit is virus dependent and ranges from 2-6 proteins.  Under certain conditions, the intensity of light scattering signal is proportional to the mass-averaged molecular weight of species in solution, which closely tracks the fraction of subunits in capsids (provided that intermediate concentrations remain small).

{\bf Single molecule techniques.} It is difficult to characterize assembly pathways with bulk techniques because most intermediates are transient. Techniques that monitor individual capsids have begun to address this limitation. For example, a Coulter-counter-like apparatus that uses resistive pulse sensing to identify the passage of individual capsids through nanopores was able to distinguish between $T{=}3$ and $T{=}4$ HBV capsids \cite{Zhou2011}. Mass spectrometry has been used to characterize key intermediates and assembly pathways for several viruses \cite{Stockley2007,Basnak2010,Uetrecht2011,Tresset2013,Pierson2014}. Fluorescent labeling of capsid proteins or RNA has enabled tracking assembly and protein-RNA association of HIV capsids in cells \cite{Baumgaertel2012,Jouvenet2011}. Borodavka et al. \cite{Borodavka2012} used single molecule fluorescence correlation spectroscopy to monitor the hydrodynamic radii of assembling nucleocapsid complexes (section~\ref{sec:NAspecific}).

{\bf Theoretical models.} Zlotnick and coworkers \cite{Zlotnick1994,Zlotnick1999,Endres2002} developed an approach to describe capsid assembly kinetics with a system of rate equations for the time evolution of concentrations of intermediates. Their equations are analogous to the classic Becker-D\"oring rate equations for cluster concentrations in a system undergoing crystallization \cite{Becker1935}, except that the capsids terminate at a finite size. The equations are made tractable by assuming one or a few structures for each intermediate size. Continuum-level descriptions of assembly dynamics (with further simplifications) have also been developed \cite{Schoot2007,Morozov2009}. The assumption of one structure per intermediate size can be relaxed by enumerating pathways \cite{Moisant2010}, or in an alternative approach \cite{Zhang2006,Keef2006,Hemberg2006,Dykeman2013a,Smith2014}, pathways consistent with a Master equation are stochastically sampled using the BKL or Gillespie algorithm \cite{Bortz1975,Gillespie1977}.

{\bf Particle-based dynamics simulations.}  The approaches described in the previous paragraph must pre-assume the state space (i.e. the possible structures of partial capsid intermediates). This limitation can be relaxed by performing simulations which explicitly track the dynamics of subunit positions and orientations using molecular dynamics, Brownian dynamics, or other equations of motion. Several groups have developed coarse-grained models for subunits, which have excluded-volume geometry and orientation-dependent attractions designed such that the lowest energy structure is a shell with icosahedral symmetry (e.g. \cite{Schwartz1998,Hagan2006,Hicks2006,Nguyen2007,Wilber2007,Nguyen2009,Johnston2010,Rapaport1999,Rapaport2008,Elrad2010,Mahalik2012,Rapaport2012}).
A recent approach uses particle-based simulations to systematically derive Markov state models, which can then be simulated using methods from the previous paragraph \cite{Perkett2014}.

\section{Empty capsid assembly }
\label{sec:empty}
We begin by analyzing the formation process of an empty capsid. While this process is most relevant to viruses that first form empty procapsids during assembly, it also provides a useful starting point to understand co-assembly with nucleic acids, lipid membranes, or scaffolding proteins.  Although we focus on icosahedral capsids, we note that many viruses have non-icosahedral capsids, and that the capsid proteins of some icosahedral viruses can form other structures including sheets, tubes, and multi-layered shells depending on solution conditions \cite{Lavelle2009}.

\subsection{Thermodynamics of assembly}
\label{sec:LMA}
We consider the thermodynamics for a system of identical protein subunits that can assemble into empty $T{=}1$ capsids. To simplify the presentation, we assume that there is one dominant intermediate species for each number of subunits $n$. Minimizing the total free energy under the constraint of fixed total subunit concentration $\rhoTot = \sum_{n=1}^N n c_n$ results in the law of mass action for the equilibrium concentration of each species $c_n$ \cite{Zlotnick1994, Safran1994,Bruinsma2003,Hagan2014}:
\begin{align}
c_n v_0= \left(c_1 v_0\right)^n \exp\left(-\Gcap_n/\kt\right)
\label{eq:rhon},
\end{align}
with with $v_0$ the standard state volume and $\kt$ the thermal energy.  Here $\Gcap_n$ is the free energy due to subunit-subunit interactions for intermediate $n$.  Zlotnick developed a class of models in which the interaction free energy is proportional to the number of subunit-subunit contacts:  $\Gcap_n = \gb C_n - T S_n$ with $C_n$ the number of subunit-subunit contacts in an intermediate, $\gb$ the subunit-subunit binding free energy, and $S_n$ a symmetry factor \cite{Zlotnick1994,Endres2002}.

Under most conditions at equilibrium, almost all of the subunits are found in complete capsids or as free subunits \cite{Zlotnick1994,Hagan2014}. This prediction arises from virtually any model for assembly of finite-size structures (e.g. capsids or micelles) in which the interaction free energy $\Gcap_n$ is minimum for one structure ($n=N$) and the total subunit concentration is conserved \cite{Safran1994}.  Under these conditions Eq.\eqref{eq:rhon} can be simplified by neglecting all intermediates except free subunits or complete capsids, so that $\rhoTot = c_1 + N c_N$ with $N$ the number of subunits in a complete capsid (i.e. a two-state approximation). Then, in the limit $N\gg1$ the fraction of subunits in capsids, $\fc = N c_N/\rhoTot$, is given by \cite{Schoot2007,Hagan2014}
\begin{align}
\fc & \approx \left(\frac{\rhoTot}{\rhoStar}\right)^N \ll 1 \qquad & \mbox{ for } \rhoTot\ll\rhoStar \nonumber \\
& \approx 1- \frac{\rhoStar}{\rhoTot} \qquad & \mbox{ for } \rhoTot\gg\rhoStar
\label{eq:fcAsympt}
\end{align}
with the `pseudo-critical' subunit concentration $\rhoStar\approx v_0^{-1}\exp\left( \Gcap_N/N\kt\right)$ below which there is no assembly.

Zlotnick and coworkers have shown that the assembly of HBV \cite{Ceres2002} can be captured by Eq.\eqref{eq:fcAsympt} using the subunit-subunit binding free energy  $\gb$ as a fit parameter. Their data shows that productive assembly requires weak binding free energies, on the order of $\gb=4$ kcal/mol ($6.7\kt$). The requirement for weak interactions appears to be quite general, for reasons discussed in section \ref{sec:emptyCapsidProducts} (see also Ref.\cite{Whitelam2014} in this issue).

\subsection{Assembly driving forces}
\label{sec:drivingForces}
Eq.~\eqref{eq:rhon} reflects the fact that formation of an ordered capsid reduces the translational entropy of its constituent subunits, and thus must be driven by favorable interactions that overcome this penalty.  These interactions (and changes in subunit rotational entropy) are described by the factor $\Gcap$. In many cases assembly is primarily driven by hydrophobic interactions, attenuated by electrostatics \cite{Alamo2005,Kegel2006} with directional specificity imposed by electrostatic, van de Waals, and hydrogen bonding interactions. These interactions are short-ranged under assembly conditions, with scales ranging from a few angstroms (Van der Waals interactions and hydrogen bonds) to $0.5-1$ nm for hydrophobic interactions \cite{Chandler2005}. Similarly, electrostatic interactions are screened on the scale of the Debye length $\lambda_\text{D}$, which is about 1 nm at physiological ionic strength (150 mM) and decreases with ionic strength $I$ according to $\lambda_\text{D} \approx 0.3/I^{1/2}$ with $\lambda_\text{D}$ in nm and $I$ in molar units.

\subsection{Empty capsid assembly mechanism}
\label{sec:emptyCapsidKinetics}
\begin{figure}[bt!]
\begin{center}
\epsfxsize=0.99\textwidth\epsfbox{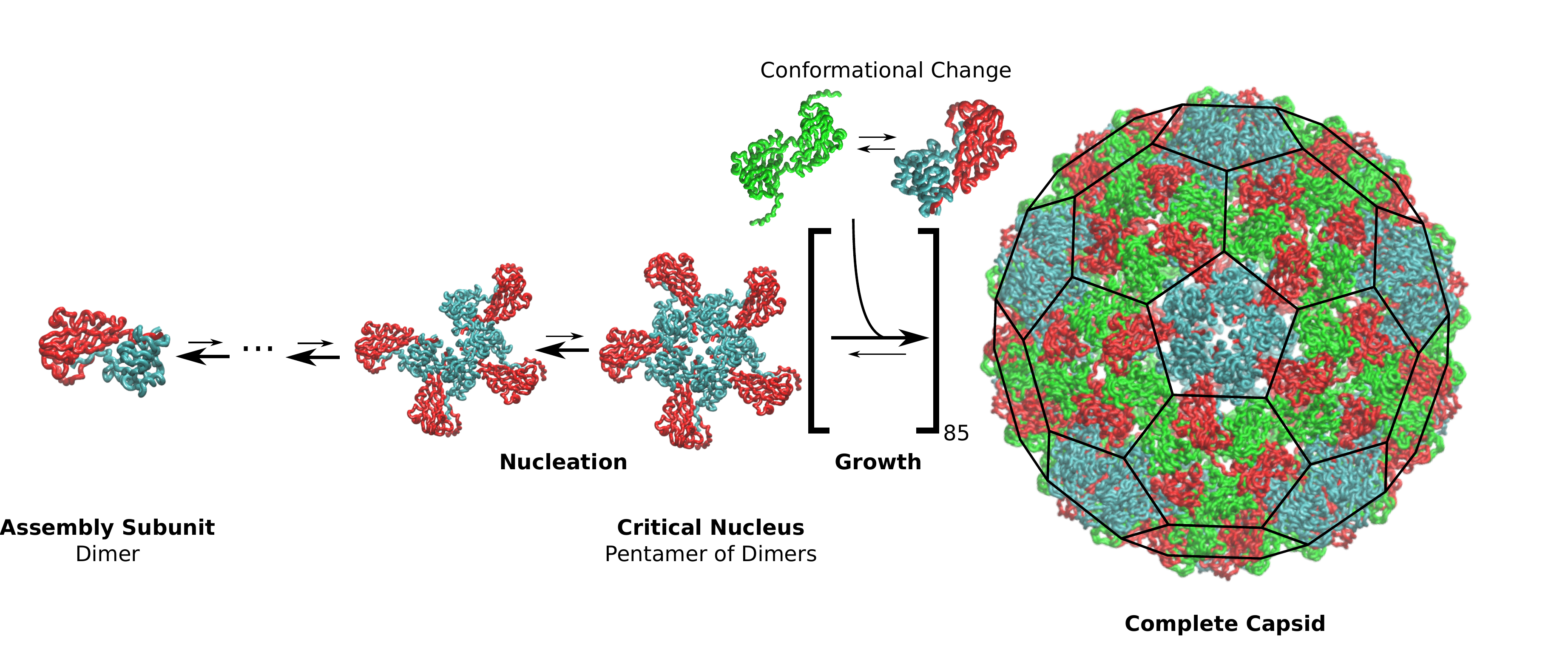}
\caption{ Schematic of the assembly mechanism for cowpea chlorotic mottle virus (CCMV) \cite{Zlotnick2000}. In the nucleation phase, addition of capsid protein dimers is unfavorable until reaching the critical nucleus. Subsequent additions (the growth phase) are relatively favorable, though still reversible, until the capsid is completed.  Subunits must interconvert between different quasi-equivalent conformations to assemble the $T{=}3$ icosahedral geometry (Fig.~\ref{fig:tNumber}); different conformations are distinguished by color. The diameter of the complete CCMV capsid is 28 nm.}
\label{fig:nucleationAndGrowth}
\end{center}
\end{figure}

As first suggested by Prevelige \cite{Prevelige1993}, empty capsids assemble by a `nucleation-and-growth' mechanism, in which a critical nucleus forms followed by a growth phase in which one or a few subunits add sequentially until the capsid is completed (Fig.~\ref{fig:nucleationAndGrowth}). The critical nucleus is defined as the smallest intermediate
 which has a greater than 50\% probability of growing to a complete capsid before disassembling. Smaller intermediates are transient and thus formation of the critical nucleus is a rare event, with a timescale $\tnuc \sim c_1^\nnuc$ with $\nnuc$ the nucleus size \cite{Endres2002,Hagan2010}.  In contrast, intermediates in the growth phase are relatively stable; thus, successive additions of subunits or small oligomers are independent and the timescale for a capsid to complete the growth phase has a low-order dependence on the free subunit concentration \cite{Hagan2010,Hagan2014}.

Due to the geometry of an icosahedral shell, the first few intermediates have relatively few subunit-subunit contacts and are thus relatively unstable. The critical nucleus often corresponds to a small polygon (Fig.~\ref{fig:nucleationAndGrowth}) whose geometry maximizes the number of interactions; furthermore, subunit conformation changes may provide additional stabilization upon polygon formation\footnote{In general there can be an ensemble of critical nuclei, whose members depend on solution conditions and protein subunit concentration.}. As the subunit-subunit binding free energy or the free subunit concentration decreases, small intermediates become less stable and the critical nucleus size increases \cite{Zandi2006}. Therefore, as subunit supersaturation decreases over the course of an assembly reaction, the critical nucleus size increases, asymptotically approaching a half capsid \cite{Hagan2014}.

{\bf Disassembly.} Similar considerations apply to capsid disassembly.  The first few subunits to disassemble must break many contacts, leading to a large activation barrier. There is therefore a pronounced hysteresis between assembly and disassembly at a given set of conditions \cite{Singh2003,Hagan2006}.  This condition allows capsids to be highly metastable even at infinite dilution \cite{Uetrecht2010a}, which is an important feature given that they must eventually leave their host cell to infect another. Some capsids undergo post-assembly maturation processes which further increase their stability.

\subsection{Empty capsid assembly kinetics and products}
\label{sec:emptyCapsidProducts}

\begin{figure} [bt!]
\begin{center}
\epsfig{file=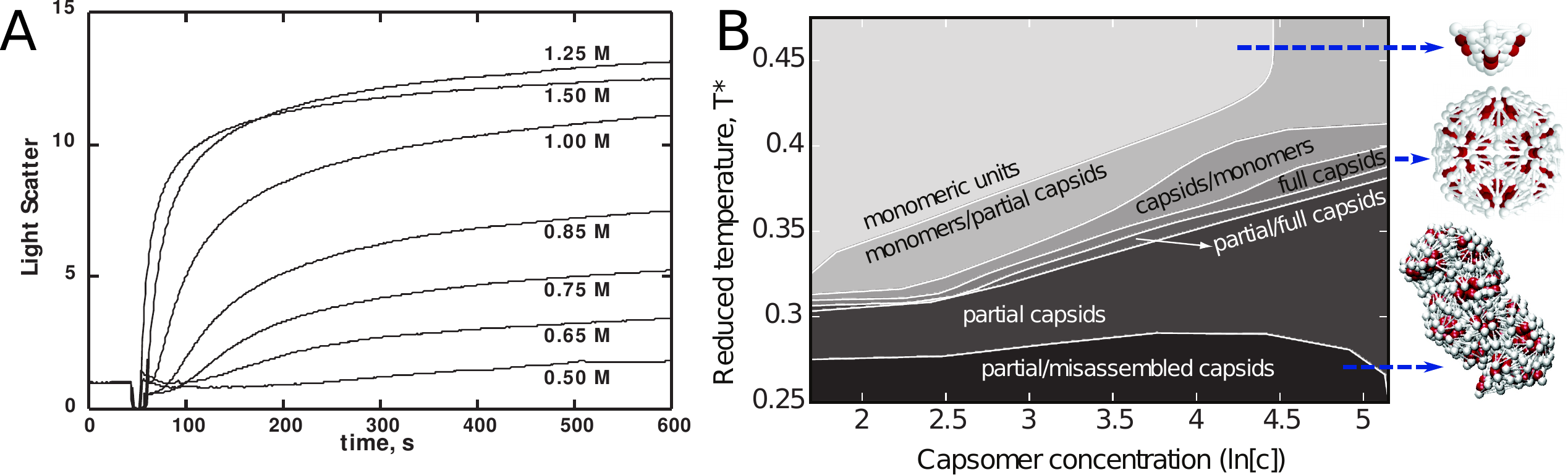,width=.99\textwidth}
 \caption{{\bf (A)} Light scattering  measured as a function of time for 5 $\mu$M dimer of HBV capsid protein at indicated ionic strengths. The image  is reprinted with permission from Ref. \cite{Zlotnick1999} Copyright (1999) American Chemical Society. {\bf (B)} Assembly products at long times for a 20-subunit icosahedral shell as a function of temperature (i.e. inverse of interaction strength) and particle concentration. Representative structures for several regions are shown on the right.  Figure adapted with permission from  Ref.~\cite{Nguyen2007}, Copyright (2007) American Chemical Society. }
  \label{fig:Nguyen}
\end{center}
\end{figure}

Capsid assembly kinetics, whether measured by experiments \cite{Prevelige1993,Zlotnick1999,Zlotnick2000,Casini2004,Chen2008,Berthet-Colominas1987,Kler2012,Kler2013} or calculated from theoretical or computational models \cite{Zlotnick1994,Zlotnick1999,Endres2002,Hagan2010,Chen2008,Kler2012,Smith2014}, are sigmoidal (Fig.~\ref{fig:Nguyen}A). There is an initial lag phase during which capsid intermediates form, followed by rapid capsid production, and then an asymptotic approach to equilibrium during which assembly slows as nucleation barriers rise due to depletion of free subunits. Increasing the subunit concentration $\rhoTot$ or the strength of inter-subunit interactions $\gb$ (typically by decreasing pH or increasing salt concentration) initially leads to more rapid assembly. However, while thermodynamics (Eq.~\ref{eq:fcAsympt}) indicates that the yield of well-formed capsids monotonically increases with $\gb$ and $\rhoTot$, the yield at long but finite times is nonmonotonic with variation of these parameters (see the highest ionic strength in Fig.~\ref{fig:Nguyen}A) due to kinetic traps.

These modeling and experimental studies show that there is a trade-off between interaction specificity and kinetic accessibility --- more specific interactions increase selectivity of assembly for the target structure but decrease assembly rates due to decreased kinetic cross-sections \cite{Whitelam2009,Whitelam2014}.  The outcome of this competition between thermodynamics and kinetics has been summarized by mapping `kinetic phase diagrams' (Fig.~\ref{fig:Nguyen}B)  which describe the predominant assembly outcome at long but finite times as a function of subunit concentration, interaction strength, and degree of interaction specificity.  For a given interaction specificity, the kinetic phase diagram can be classified into five regimes.

\textit{(i) No assembly at equilibrium:} For weak interactions or low subunit concentrations, such that $\rhoTot<\rhoStar$ (Eq.\eqref{eq:fcAsympt}), assembly is unfavorable at equilibrium.  \textit{(ii) Prohibitive nucleation barriers:} As interactions or subunit concentrations increase to $\rhoTot \gtrsim \rhoStar$, assembly is favorable, but does not occur on experimentally relevant timescales due to large nucleation barriers. \textit{(iii) Productive assembly:} Further increasing interactions or subunit concentrations leads to moderate nucleation barriers and large yields of well-formed capsids on relevant timescales (which can range from seconds to hours for empty capsids).  Finally, stronger-than-optimal interactions lead to suppressed yields due to two forms of kinetic traps. \textit{(iv) Free subunit starvation kinetic trap:}  When nucleation is fast compared to growth, too many capsids nucleate at early times and free subunits or small intermediates are depleted before a significant number of capsids finish assembling \cite{Zlotnick1999,Zlotnick2000,Endres2002,Hagan2006,Nguyen2007,Rapaport2008}. This condition occurs when the timescale required for capsids to complete the growth phase exceeds the typical nucleation timescale \cite{Hagan2010,Hagan2014}.

\textit{(v) Malformed capsids:}  Under sufficiently strong interactions, subunits with imperfect orientations are trapped into growing clusters by subsequent subunit additions, leading to either defective closed shells that lack icosahedral symmetry or open, spiral structures in simulations \cite{Schwartz1998,Hagan2006,Hicks2006,Nguyen2007,Nguyen2009} and experiments \cite{Sorger1986,Stray2005,Parent2007a}. The presence of these two forms of kinetic traps (\textit{iv} and \textit{v}) explain the experimental \cite{Ceres2002,Zlotnick2003} and computational \cite{Hagan2006,Rapaport2008} observation that weak interactions are required for productive capsid assembly.  These kinetic traps lead to similar constraints on interactions in other forms of assembly such as crystallization \cite{Whitelam2014}.

\section{Capsid assembly around nucleic acids and other polyelectrolytes}
\label{sec:cargo}
This section focuses on viruses for which the capsid assembles spontaneously around the viral genome during infection. This category includes most ssRNA viruses and the Hepadnaviridae (e.g. HBV), and Spumaviridae, which assemble around ssRNA pregenomes that then undergo reverse transcription to yield dsDNA within the virions.

Electrostatic interactions between positive charges on capsid proteins and negative charges on RNA provide an important thermodynamic driving force for this process. For example, the capsid proteins of many negative-stranded RNA viruses bind RNA via a positively-charged cleft \cite{Ruigrok2011}. For many positive-stranded RNA viruses, the capsid proteins bind RNA via flexible terminal domains rich in basic amino acids, called arginine rich motifs (ARMs)\cite{Speir1995}. Specific RNA sequences or chemistry are not essential for assembly, as demonstrated by early in vitro experiments in which ssRNA capsid proteins assembled around heterologous nucleic acids and even polyvinylsulfate \cite{Hohn1969,Bancroft1969}, and more recent experiments in which  capsid proteins assembled around various negatively charged substrates (e.g. \cite{Bancroft1969,Hohn1969,Chen2005,Sun2007,Hu2008d,Sikkema2007,Brasch2012,Kostiainen2011,Goicochea2007,Loo2007,Kwak2010,Chang2008,Malyutin2013,Cheng2013}).  We begin this section by describing what these experiments and theoretical models have revealed about how assembly depends on the physical characteristics of RNA or other polyelectrolytes, such as charge, size, and structure. Due to space limitations, we do not discuss studies on assembly around non-polymeric cores (see Refs.~\cite{Hagan2014,Siber2012}). We then discuss mechanisms by which virus-specific interactions can enhance co-assembly and enable selective packaging of the viral genome.

\subsection{Thermodynamics of assembly around a cargo}
\label{sec:cargoThermo}
  We consider a solution of capsid protein subunits and cores (e.g. RNA molecules) with respective total concentrations of $\rhoTot$ and $\coreTot$.  We define a stoichiometric ratio as the ratio of available cores to the maximum number of capsids which can be assembled, $r=N \coreTot/\rhoTot$. Extending  Eq.~\ref{eq:rhon} to include interior cores results in two laws of mass action\cite{Zandi2009,Hagan2009,Hagan2014}:
  \begin{align}
c_n v_0&=\left(c_1 v_0\right)^n\exp[-\Gcap_n/\kt]  \label{eq:empty} \\
x_n v_0 &=x_{0} v_0 \left(c_1 v_0\right)^{n}\exp[-\Gcore_n/\kt]
\label{eq:Zcore}
\end{align}
with $x_0$ the concentration of empty cores. Eqs. ~\ref{eq:empty} and \ref{eq:Zcore} describe assembly of empty and core-containing capsids with respective interaction free energies $\Gcap$ and $\Gcore$. The core interaction energy $\Gcore$ includes, for example in the case of an RNA core, attractive protein-RNA interactions, intramolecular electrostatic repulsions, and base-pairing interactions.
Eq.~\ref{eq:Zcore} identifies a new critical subunit concentration, which for excess capsid protein is given by $\rhoStarStar = \exp\left[ \Gcore_N/N \kt\right]/v_0$ \cite{Zandi2009}. If the net contribution of the core to assembly is favorable ($\Gcore_N < \Gcap_N$) core-assisted assembly can occur at concentrations below the threshold concentration for empty capsid assembly ($\rhoStarStar < \rhoStar$). This capability is exploited by  many ssRNA viruses, whose capsids assemble only in the presence of RNA or other polyanions at physiological conditions, thus ensuring that the genome is packaged during assembly. On the other hand,  unfavorable core contributions, such as would arise from the stiffness and electrostatic repulsions of dsDNA molecules, can direct assembly away from the capsid structure to other morphologies \cite{Zlotnick2013}.

\subsection{Optimal genome length}
\label{sec:optimalGenome}
\begin{figure} [bt!]
\begin{center}
\epsfxsize=0.8\textwidth\epsfbox{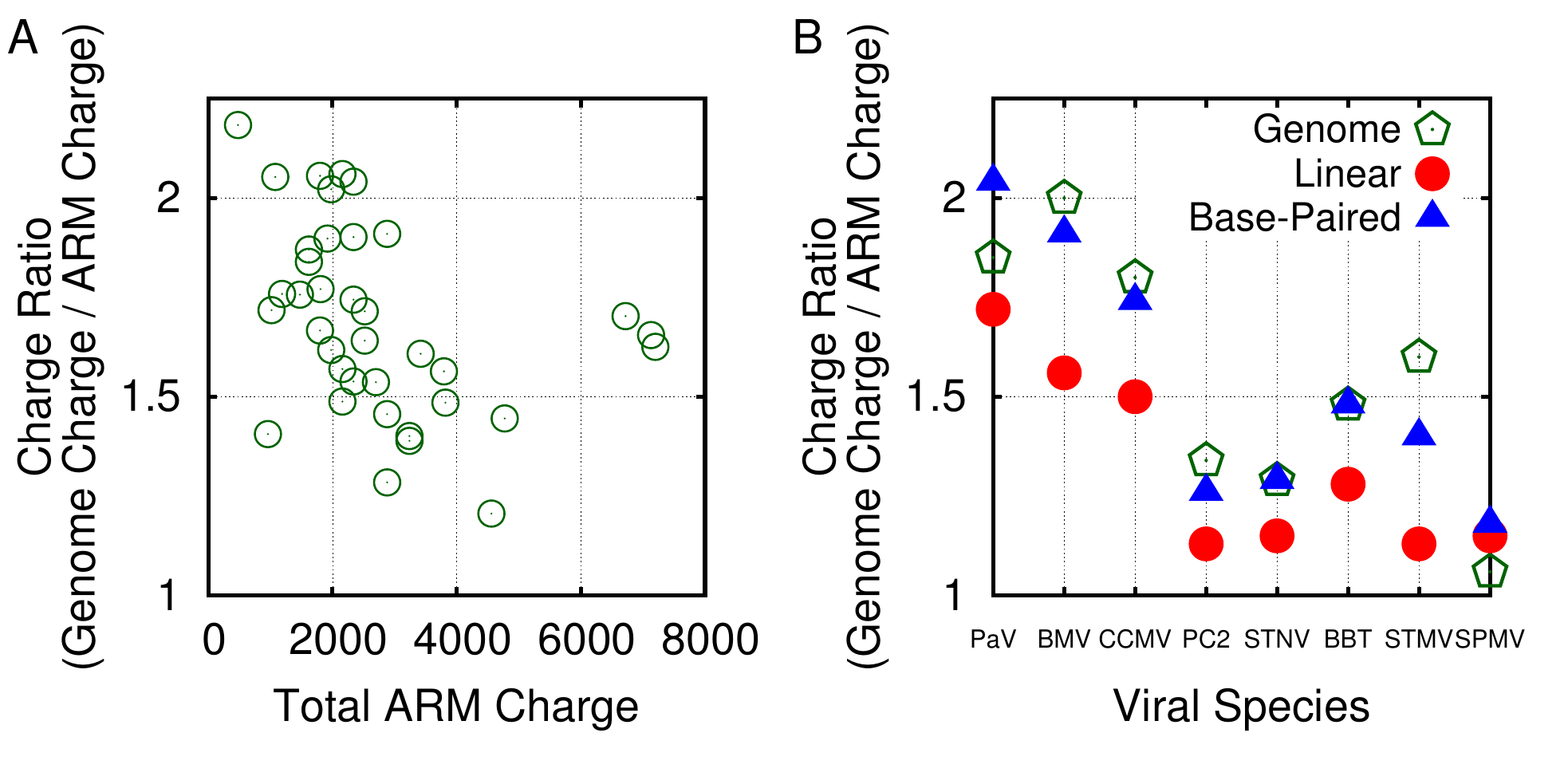}
\caption{ Relationship between genome length and capsid charge. {\bf (A)} Survey of the charge ratio, or number of nucleotides in the genome divided by total positive charge on the inner capsid surface, for ssRNA viruses.  {\bf (B)} The thermodynamic optimum charge ratio predicted from simulations \cite{Perlmutter2013} (\textcolor{blue}{$\blacktriangle$} symbols)  is compared to actual charge ratios (\textcolor{green}{\pentagon} symbols ) for several viruses. Predicted optimal charge ratios in the absence of base-pairing are also shown ({\Large\textcolor{red}{$\bullet$}} symbols).  The thermodynamic optimum charge ratio is defined  as the NA length which minimizes the free energy for encapsidating the genome divided by the positive capsid charge.}
\label{fig:chargeRatio}
\end{center}
\end{figure}

It has been proposed that viral genomes face a selective pressure to maintain a length which maximizes the stability of the nucleocapsid complex (i.e., minimizes $\Gcore_N$).  In support of the importance of nonspecific electrostatics to driving RNA encapsidation,  the total positive charge on the capsid inner surface correlates to the length of the genomic RNA for a diverse group of ssRNA viruses \cite{Belyi2006,Hu2008a} (Fig.~\ref{fig:chargeRatio}). Furthermore, changing the capsid charge alters the amount of cargo encapsidated in cells and in vitro \cite{Newman2009,Venter2009, LePogam2005, Ni2012, Porterfield2010}, although the effect of  mutations on the amounts and sequences of packaged RNA can depend on factors other than charge\cite{Kao2011,Ni2012}. Importantly, ssRNA viruses are consistently overcharged, with typical charge ratios of $\chargeratio=|\mbox{NA charge}|/|\mbox{protein charge}|$ in the range $1.5 \le \chargeratio \le 2$ (Fig.~\ref{fig:chargeRatio}). In vitro competition assays in which different species of RNAs competed for packaging demonstrated that longer RNAs (up to the viral genome length) are preferentially packaged over shorter RNAs \cite{Comas-Garcia2012}, indicating that overcharged genomes are optimal for packaging even in the absence of cell-specific factors.

Motivated by these observations, researchers theoretically and computationally calculated how the free energy $\Fe(L)$ to encapsulate a linear polyelectrolyte  varies with its length $L$ (reviewed in \cite{Siber2012,Hagan2014}). Several works performed self-consistent field theory calculations in which $F(L)$ is calculated from a continuum description of polymer conformational statistics coupled to the Poisson-Boltzmann equation. While Ref. \cite{Schoot2005} predicted an optimal charge ratio of $\chargeratio=2$, most subsequent calculations predicted $\chargeratio \lesssim 1$ \cite{Siber2008,Siber2012,Ni2012,Belyi2006,Ting2011}.  Ref. \cite{Belyi2006} noted that if the charge on the RNA and the peptide tails were renormalized according to counterion condensation theory \cite{Manning1969b} overcharging would be predicted; however, the condensed counterions are released by RNA-peptide association and thus the charge cannot simply be renormalized \cite{Perlmutter2013}.
Ting et al. \cite{Ting2011} found that the optimal charge ratio varies with capsid volume and the charge density on capsid protein ARMs (i.e., there is no single optimal charge ratio). Since in all cases the model predicted $\chargeratio<1$, they suggested that a Donnan potential arising from negatively charged macromolecules within cells drives overcharging.  However, the subsequent in vitro competition assays \cite{Comas-Garcia2012} demonstrated that overcharging is optimal for assembly in the absence of a Donnan potential.

Perlmutter et al. \cite{Perlmutter2013} used a coarse-grained particle-based computational model, in which ARMs were represented as flexible polyelectrolytes affixed to capsid subunits, to determine the optimal lengths of encapsulated polyelectrolytes and NAs as functions of capsid size, ARM charge, and ionic strength. The model predicted overcharging ($\chargeratio>1$) in all cases. Optimal lengths predicted by the model for several specific viruses closely matched genome lengths for those viruses (Fig.~\ref{fig:chargeRatio}B). Overcharging was found to arise because only a fraction of encapsulated polymer segments can closely interact with positive capsid charges (i.e. within a Debye length).  Consequently, packaging of multiple short polyelectrolytes will lead to reduced or no overcharging. The thermodynamic optimal lengths closely matched the lengths which optimized the yield of long but finite-time dynamical simulations, indicating a connection between the thermostability and optimal assembly of a viral particle.

{\bf  Effect of RNA base-pairing.} About 50-60\% of nucleotides undergo intramolecular base-pairing in ssRNA molecules, leading to compact, branched structures, as recently visualized using cryo-EM \cite{Gopal2012}. Although a self-consistent field theory predicted no difference between the optimal lengths for linear polyelectrolytes and compact star architectures\cite{Ting2011}, subsequent theory \cite{Erdemci-Tandogan2014} and simulations \cite{Perlmutter2013} found that branching consistent with the structures of base-paired RNA increases the optimal genome length as compared to a linear polyelectrolyte, by compensating for intramolecular charge repulsions and by favoring compact conformations (Fig.~\ref{fig:chargeRatio}B). Based on secondary structure predictions, Yoffe et al. \cite{Yoffe2008} suggested that viral RNAs tend to have more compact tertiary structures than cellular RNAs with equal numbers of nucleotides, which could favor assembly around the viral RNA.

\begin{figure}[bt!]
\begin{center}
\epsfxsize=0.99\textwidth\epsfbox{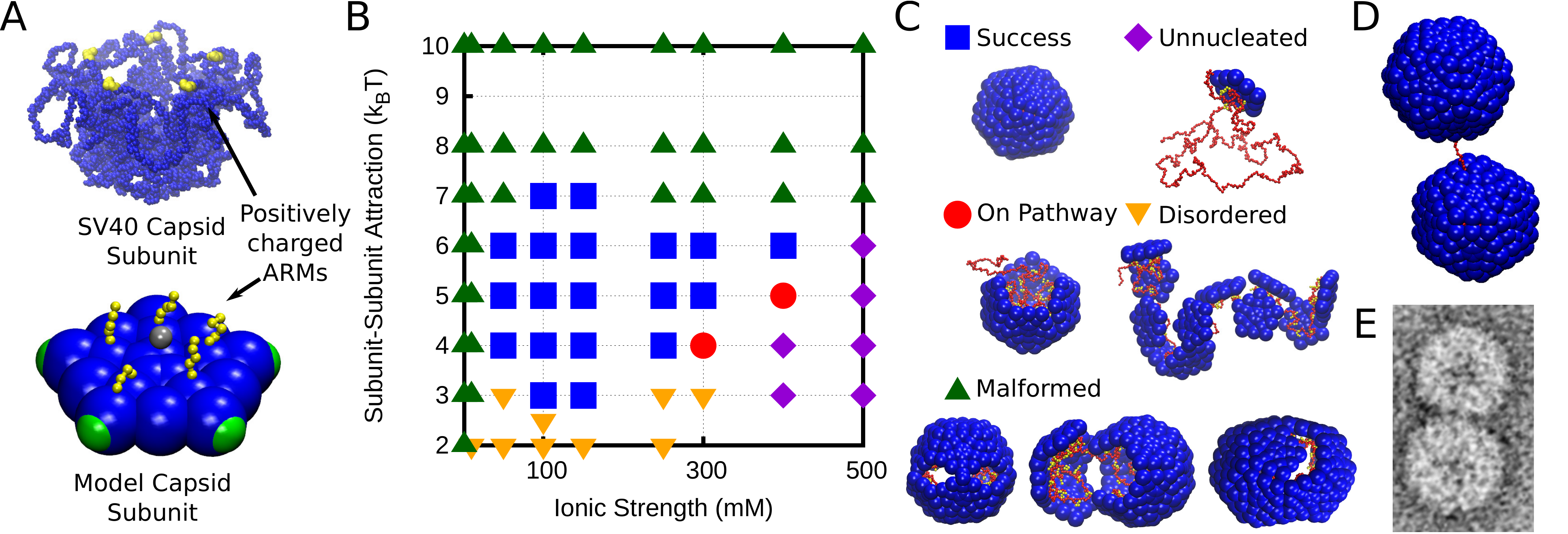}
\caption{{\bf (A)} Crystal structure of the SV40 basic assembly unit \cite{Stehle1996}, which is a homopentamer of the capsid protein capsid subunit, and a coarse-grained model pentameric subunit. The locations of the positively charged ARMs are shown in yellow (most of the ARM residues are not resolved in the crystal structure). {\bf (B)} The dominant products of assembly around a linear polyelectrolyte as a function of ionic strength and subunit-subunit interaction strength at thermodynamically optimal polyelectrolyte lengths, which vary from 350-575 depending on the ionic strength. {\bf (C)} Simulation snapshots which exemplify the dominant assembly outcomes. {\bf (D)} A doublet formed in simulations around a polyelectrolyte with 1200 segments (twice the optimal length). {\bf (E)}  A doublet assembled from CCMV capsid proteins around RNA with 6400 nucleotides (about twice the number of nucleotides encapsidated in native CCMV virions) \cite{Cadena-Nava2012}. Image provided by R. Garmann, C. Knobler and W. Gelbart.}
\label{fig:polymerOutcomes}
\end{center}
\end{figure}

\subsection{Assembly at non-optimal parameters}
\label{sec:outcomes}
The previous section showed that, for a given capsid protein and solution conditions, there is a length of RNA for which assembly is optimal. To understand the effect of perturbing parameters from these optimal values, in vitro assembly products were characterized by electron microscopy (e.g. \cite{Hu2008d,Cadena-Nava2012,Comas-Garcia2012,Garmann2014,Brasch2012}) or SAXS \cite{Kler2012,Kler2013} as functions of subunit and RNA concentrations, subunit-subunit interactions (controlled by pH, ionic strength, and protein sequence), and subunit-polymer interactions (controlled by ionic strength and protein-RNA binding domains). In addition, several groups have performed Brownian dynamics simulations in which coarse-grained triangular or pentameric subunits assemble around flexible polyelectrolytes \cite{Elrad2010,Mahalik2012,Perlmutter2013,Perlmutter2014, Zhang2013,Zhang2013a}, semiflexible polyelectrolytes \cite{ Zhang2013a}, or model NAs \cite{Perlmutter2013}. In both experiments and simulations, parameters must be carefully tuned to achieve high yields of well-formed capsids. An example simulation phase diagram illustrating some of the alternative products that form at non-optimal parameters is shown in Fig.~\ref{fig:polymerOutcomes}.

Assembly around polymers with non-optimal lengths leads to several outcomes. The first is polymorphism, or formation of capsids with different $T$-numbers (Fig.~\ref{fig:tNumber}). CCMV capsid proteins (which form native $T{=}3$ capsids) formed $T{=}3$ capsids around genomic-length RNA and pseudo-$T{=}2$-sized capsids around shorter RNAs \cite{Cadena-Nava2012,Comas-Garcia2012}.  The favored polymorph depends on RNA length, the preferred curvature of capsid protein-protein interactions (i.e. spontaneous curvature, section~\ref{sec:membraneThermo}) and stoichiometry \cite{Hu2008d,Zandi2009}.  For RNA significantly below optimal length, multiple RNAs were packaged in each capsid \cite{Cadena-Nava2012,Comas-Garcia2014}, as seen in coarse-grained dynamics simulations with short linear polyelectrolytes\cite{Zhang2013}.  In the experiments, capsid formation required equilibrium between multiple disordered protein-RNA complexes, leading to highly cooperative assembly \cite{Comas-Garcia2014}.

While longer-than-optimal polymers sometimes led to $T{=}4$-sized capsids, the more common outcome was the assembly of multiplets, or multiple distinct capsids assembled around one RNA \cite{Cadena-Nava2012} or conjugated polyelectrolyte  \cite{Brasch2012}. RNAs which were 2, 3, or 4 times the genome length lead respectively to predominantly doublets (Fig.~\ref{fig:polymerOutcomes}D), triplets, and quadruplets. An early study likely also observed doublets, although the structures could not be confirmed \cite{Hohn1969} and recently doublet dodedecadron capsids were observed for assembly of SV40 capsid proteins around certain lengths of RNA \cite{Kler2013}. Simulations independently predicted the formation of doublets around polyelectrolytes with about twice  the optimal length \cite{Kivenson2010,Elrad2010, Perlmutter2013} (Fig.~\ref{fig:polymerOutcomes}C). At lengths only slightly greater than optimal, simulations predict malformed but single capsids \cite{Elrad2010,Perlmutter2013,Mahalik2012}. However, these malformations may be difficult to resolve experimentally.

\subsection{Assembly mechanisms}
\label{sec:assembly_mechanisms}
\begin{figure} [bt!]
\begin{center}
\includegraphics[width=1.0\textwidth]{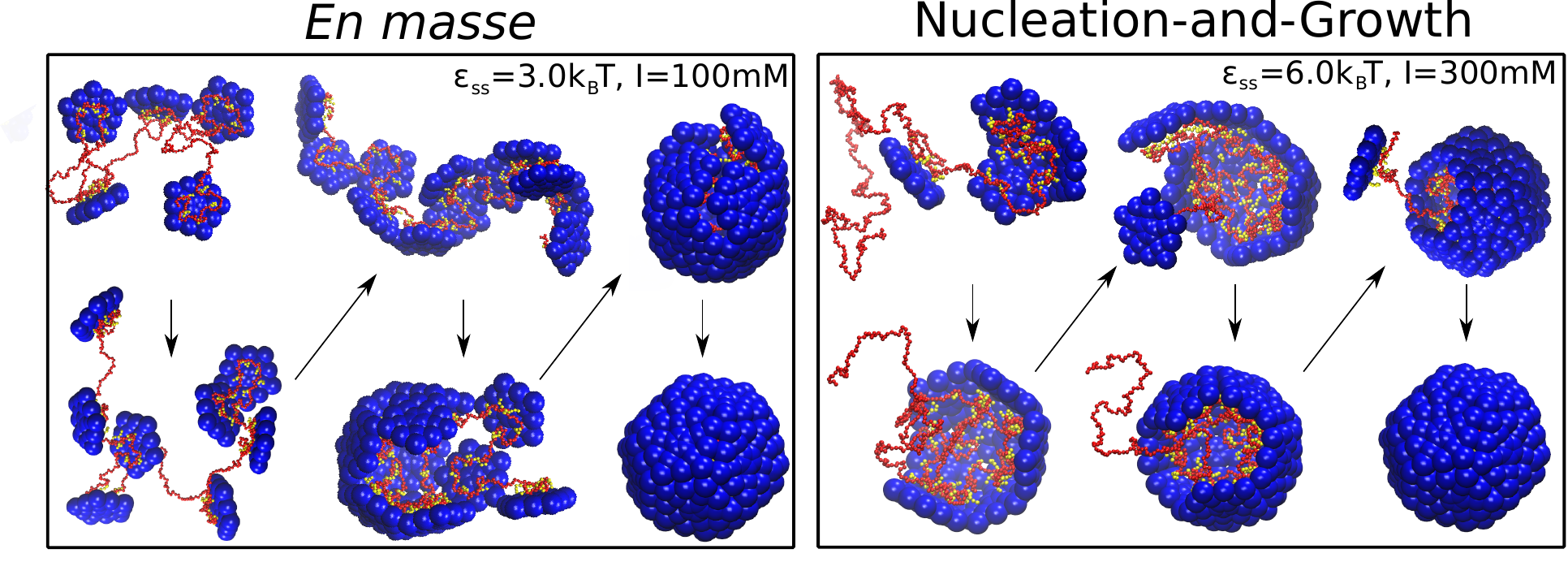}
\caption{Two mechanisms for assembly around a polyelectrolyte \cite{Perlmutter2014}.  {\bf (A)} Low ionic strength (strong subunit-polyelectrolyte interactions) and weak subunit-subunit interactions lead to the \emph{en masse} mechanism typified by disordered intermediates. {\bf (B)} High ionic strength (weak subunit-polymer interactions) and strong subunit-subunit interactions lead to the nucleation-and-growth mechanism in which an ordered nucleus forms on the polymer followed by sequential addition of subunits. }
\label{fig:polymer_mechanisms}
\end{center}
\end{figure}

Simulations \cite{Elrad2010,Perlmutter2014} show that two classes of assembly mechanisms occur around RNA or a linear polymer (Fig. \ref{fig:polymer_mechanisms}). One closely resembles the nucleation-and-growth mechanism found for empty capsid assembly, except that the polymer stabilizes protein-protein interactions and can enhance the flux of proteins to the assembling capsid \cite{Hu2007}. A small partial capsid first nucleates on the polymer, followed by a growth phase in which one or a few subunits sequentially and reversibly add to the partial capsid. In the alternative mechanism, first proposed by McPherson \cite{McPherson2005} and then Refs~\cite{Hagan2008,Devkota2009,Elrad2010}, subunits adsorb onto the polymer \emph{en masse} in a disordered fashion and then cooperatively rearrange to form an ordered capsid. Simulations predict that the assembly mechanism can be tuned by solution conditions and capsid protein-protein interactions\cite{Perlmutter2014}. The nucleation-and-growth mechanism is favored by weak protein-polymer association (high salt concentration) and strong protein-protein interactions (typically low pH \cite{Zlotnick2013}), while the \emph{en masse} mechanism arises for lower salt and weaker protein-protein interactions.

Observations in vitro suggest that both mechanisms are viable. Time-resolved SAXS experiments monitoring assembly of SV40 capsid proteins assembling around ssRNA produced scattering profiles which could be decomposed into profiles corresponding to unassembled components (RNA + protein subunits) and complete capsids \cite{Kler2012}.  The absence of detectable intermediates suggested that assembly follows the nucleation-and-growth mechanism. In support of this conclusion, simulations \cite{Perlmutter2014} showed that the disordered intermediates arising in \emph{en masse} pathways lead to measurably different SAXS profiles, whereas profiles from nucleation-and-growth trajectories are consistent with the experimental observations. Other observations suggest that virus-like particles can assemble through the \textit{en masse} mechanism. Refs. \cite{Cadena-Nava2012,Garmann2014,Garmann2014a} found that \textit{in vitro} assembly CCMV assembly was most productive when performed in two steps. First, at low salt (strong protein-RNA interactions) and neutral pH (weak protein-protein interactions) the proteins undergo extensive but disordered adsorption onto RNA. Subsequently, pH is reduced to enhance protein-protein binding, leading to the formation of ordered capsids \cite{Garmann2014}. Similarly, a recent observation of capsid protein assembly around charge-functionalized nanoparticles found that assembly initially proceeded through nonspecific aggregation of proteins and nanoparticles, followed by the gradual extrusion of complete capsids formed around nanoparticles \cite{Malyutin2013}.

The CCMV experiments \cite{Cadena-Nava2012,Garmann2014,Garmann2014a} found that complete encapsidation of all RNA present in solution requires a significant excess of capsid protein, such that the positive charges in protein ARMs balance the negative RNA charge (recall that in the complete capsid the negative RNA charge significantly exceeds the positive ARM charge, section~\ref{sec:optimalGenome}). This criteria occurs because the disordered protein-RNA complexes occurring during the first step of assembly are charge-balanced.  During capsid formation (the second step), excess proteins are displaced to the exterior, where their positive ARM charges interact with negative residues on the outer surface of the capsid \cite{Garmann2014a}.

\subsection{Sequence-specific contributions to assembly and selective genome packaging}
\label{sec:NAspecific}

To be infectious, a virion must assemble specifically around the viral genome amidst a panoply of cellular RNA molecules. Many viruses achieve high specificity; for example, a recent quantitative analysis found that flock house virus particles are 99\% selective for the viral RNA \cite{Routh2012}. Structure- and sequence-specific RNA-protein interactions may be a widespread  mechanism of achieving specificity by promoting assembly around the viral genome (although not all viruses are selective for their genomic RNA in vitro \cite{Porterfield2010,Comas-Garcia2012}) , suggesting the importance of cell-specific factors).

Several studies suggest that the specific folded structure of the genomic RNA may enhance assembly. Based on the crystal structure of STMV, which shows 30 ds helical segments interacting with the capsid inner surface \cite{Larson1993,Larson1998}, McPherson and coworkers \cite{Larson2001} proposed that during assembly the STMV genome forms a conformation comprising linearly connected stem loops which sequentially bind capsid proteins. Using the crystal structure as constraints, Schroeder et al. \cite{Schroeder2011} combined chemical probing and computational methods to predict an RNA secondary structure containing 30 stem-loops. Simulation of the complete STMV capsid with atomic resolution demonstrated that this secondary structure is consistent with the crystal structure \cite{Zeng2012}. More recently, two studies \cite{Archer2013,Athavale2013} used the chemical probe method SHAPE to characterize unencapsidated STMV RNA.  These analyses were not restricted to stem-loops, and found secondary structures that differed significantly from the encapsidated, stem-loop structure. Interestingly though, the primary probing data is similar for the encapsidated and unencapsidated RNAs, suggesting that the same nucleic acids are base-paired in both cases.

Extensive evidence shows that packaging signals, or short RNA sequences that are specifically bound by capsid proteins, play significant roles in controlling assembly pathways for some viruses (reviewed in  \cite{Rao2006,Stockley2013a}). Packaging signals have been identified for several viruses including HIV \cite{Pappalardo1998,Lu2011,D'Souza2005}, MS2, and STNV \cite{Bunka2011,Dykeman2011,Dykeman2013}, and a number of plant viruses \cite{Rao2006}. Combining identified packaging signals  with geometric constraints derived from electron density maps of MS2 capsids led to a structure of the encapsidated genome \cite{Dykeman2011}. Earlier work using mass spectrometry \cite{Stockley2007}, coarse-grained simulations \cite{ElSawy2010}, and kinetic models \cite{Morton2010} suggested that RNA binding drives a conformational switch in the MS2 capsid protein and identified two dominant pathways for MS2 assembly. Using Gillespie algorithm simulations (section~\ref{sec:measurements}), Dykeman et al. showed that packaging signals could enhance yields of capsids assembling around RNA in comparison to a polymer cargo with uniform interactions \cite{Dykeman2013a} and that specificity for the genomic RNA can be enhanced by time-dependent capsid protein production rates in bacteria \cite{Dykeman2014}.

A striking observation supporting an active, sequence-specific role of the genome was made by Borodavka et al. \cite{Borodavka2012}, who used single molecule fluorescence correlation spectroscopy (smFCS) to monitor the hydrodynamic radii $\Rh$ of nucleocapsid complexes. Assembly around genomic RNAs was characterized by either constant $\Rh$ or, in some trajectories, a collapsed complex followed by gradual increase to the size of an assembled capsid. In contrast, assembly around heterologous RNA led to  an increase in $\Rh$ before eventually decreasing to the size of the capsid. The different assembly pathways were attributed to the presence of packaging signals in the genomic RNAs. The collapsed structures are reminiscent of a previous observation \cite{Johnson2004}, in which incubation of CCMV RNA with sub-stoichiometric concentrations of capsid proteins led to a compact nucleocapsid complex that triggered rapid assembly upon introduction of additional capsid proteins.

Finally, we emphasize that sequence-specific protein-RNA interactions are not the only mechanism that drives selective genome packaging in vivo; other factors include, e.g. coordination of assembly with RNA replication \cite{Annamalai2008,Rao2006}.  As evidence for this, in (in vitro) competition assays  HBV capsid proteins show no preference for genomic RNA over heterologous RNA with equal length \cite{Porterfield2010} and CCMV capsid proteins preferentially assemble around  BMV RNA over the genomic CCMV RNA \cite{Comas-Garcia2012}. Since BMV and CCMV RNAs are of similar length, it was proposed that RNA tertiary structure could drive preferential encapsidation of BMV RNA.

\section{Capsid assembly on membranes}
\label{sec:membrane}
In this section we consider mechanisms by which the proteins of enveloped viruses assemble on lipid bilayers to drive budding. The passage of nanoscale particles through membranes is an extremely broad topic;  we focus on viral budding driven by protein assembly.

\begin{figure} [bt!]
\begin{center}
\epsfxsize=0.99\textwidth\epsfbox{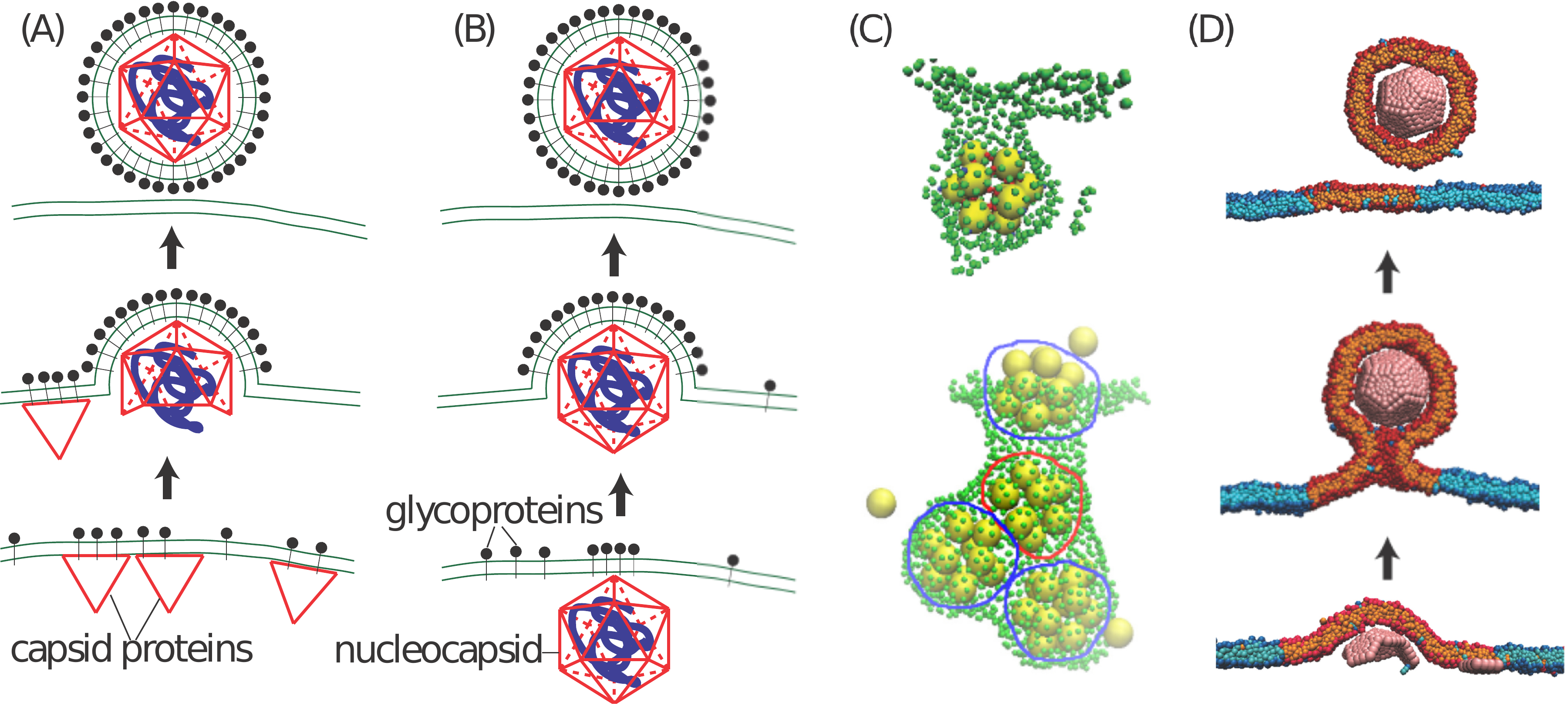}
\caption{Viral budding pathways. {\bf (A),(B)}  Schematic of the two classes of budding pathways for enveloped viruses. {\bf (A)} Assembly of capsid proteins (CPs, red) drives budding and recruitment of glycoproteins (e.g. type C retroviruses). {\bf (B)} Glycoproteins (GPs, black) drive budding of a pre-assembled nucleocapsid core (e.g. alphaviruses). {\bf (C)} Snapshots from simulations in which patchy sphere icosahedrons assemble on and bud from a triangulated membrane. The top image is reprinted with permission from Ref. \cite{Matthews2012} Copyright (2012) American Institute of Physics. The bottom image is reprinted from Ref. \cite{Matthews2013a}. {\bf (D)} Model subunits assembling in and budding from a membrane microdomain \cite{Ruiz-Herrero2014}. }
\label{fig:buddingSchematic}
\end{center}
\end{figure}

Enveloped viruses can be divided into two groups based on how they acquire their lipid membrane envelope.  For the first group, which includes influenza and type C retroviruses (e.g. HIV), the nucleocapsid core assembles on the membrane concomitant with budding (Fig.~\ref{fig:buddingSchematic}A). Capsid protein (CP) binding to membranes is driven by electrostatic interactions between positive residues on capsid proteins (e.g. the MA domain of HIV GAG) and negative charges in lipid head groups, and/or insertion of CP hydrophobic moieties (e.g. the myristoyl domain on GAG) into the bilayer \cite{Hurley2010,Sundquist2012}. In the second group, a core assembles in the cytoplasm prior to envelopment (reviewed in \cite{Sundquist2012,Welsch2007,Hurley2010}). In many families from this group envelopment of the core is driven by assembly of viral transmembrane glycoproteins (GPs) which form an outer shell around the core (Fig.~\ref{fig:buddingSchematic}B), as shown by the fact that expression of GPs alone can drive budding \cite{Garoff2004}. Thus, in both groups membrane deformation is driven at least in part by reversible protein-protein and protein-lipid interactions.  However, some viruses are assisted by cellular factors that create or support membrane curvature \cite{McMahon2005,Doherty2009,Hurley2010} or cytoskeletal machinery that actively drives budding (e.g. \cite{Taylor2011,Gladnikoff2009,Welsch2007,Balasubramaniam2011a}). Furthermore, separation of budded viral particles from the host membrane (scission) is driven by cell membrane remodeling machinery \cite{Rossman2013}.

Enveloped viruses have been the subject of extensive structural studies and in vivo investigations  \cite{Sundquist2012,Welsch2007,Hurley2010}, with budding of individual capsids from cells monitored using fluorescently labeled CPs and RNA \cite{Baumgaertel2012,Jouvenet2011}. However, there are currently no in vitro systems in which enveloped viruses assemble and bud, and information about assembly mechanisms and budding kinetics is limited in comparison to assembly in solution.

\subsection{Thermodynamics of membrane-associated assembly.}
\label{sec:membraneThermo}
The thermodynamics of assembly on a membrane can be obtained by extending the analysis in section~\ref{sec:LMA} to include membrane bending energy and subunit-membrane interactions. The free energy associated with membrane deformations on scales large in comparison to the 5 nm width of a lipid bilayer can be analyzed according to continuum elasticity (the Helfrich free energy) \cite{Helfrich1973,Safran1994}:
\begin{align}
\label{eq:Ebend}
\Ebend & =  \int dA \left[ \frac{\kappa}{2} (H - H_0)^2 + \kappa_\text{G} K \right] + \int \gamma dA  
\end{align}

with $H = (1/R_1 + 1/R_2)$ as the mean curvature, $K= 1/(R_1 R_2)$ as the Gaussian curvature, and $R_1$ and $R_2$ as the principal radii of curvature at each point on the membrane neutral surface. The remaining parameters are material properties: $H_0$ is the membrane spontaneous curvature, $\kappa$  and $\kappa_\text{G}$ are respectively the bending modulus and Gaussian modulus, $H_0$ is the membrane spontaneous curvature, and $\gamma$ is the surface tension.

Because scission is actively driven by cell machinery, it is instructive to calculate the membrane bending energy for a completely budded viral particle just before scission (i.e., the bending energy of a vesicle).  Since the total Gaussian curvature is constant for fixed topology, the corresponding term in Eq.\eqref{eq:Ebend} can be neglected (for uniform $\kappa_\text{G}$).  Assuming a tensionless membrane, that the membrane envelope is roughly spherical, and that $H_0=0$ (spontaneous curvature induced by protein binding will be considered next), Eq.\eqref{eq:Ebend} gives a fixed cost for any radius $\Ebend(\mbox{sphere}) = 8 \pi \kappa$.  Lipid bilayer bending moduli are typically in the range $10 < \kappa < 30$ $\kt$, giving a membrane bending energy cost of 250-750 $\kt$ for any size virus.

This bending energy must be compensated  by CP-membrane, GP-GP, GP-CP, and/or CP-CP interactions, which conspire to drive membrane curvature either through the spontaneous curvature of the protein-protein interaction geometry or directly through protein-membrane interactions. Including these effects and the bending energy Eq.\eqref{eq:Ebend} in the law of mass action (section~\ref{sec:LMA}) predicts that assembly proceeds above a threshold subunit concentration given by:
\begin{align}
\rhoStar v_0 \approx & \exp\left(G_N/N \kt\right) \nonumber \\
G_N/N = & \Gcap_N/N + \gsm + 8 \pi \kappa/N
\label{eq:fcMembrane}
\end{align}
with $\gsm$ accounting for the subunit-membrane interactions (per subunit) in a completely assembled and enveloped capsid. The second line of Eq.\eqref{eq:fcMembrane} emphasizes that, because the membrane bending energy is independent of capsid size, its effect is inversely proportional to the number of subunits. Thus, larger capsids are favored; perhaps relatedly, the smallest enveloped viruses have 240 proteins ($T{=}4$). If $|\gsm|> 8\pi \kappa/N$, assembly can proceed on the membrane for subunit concentrations at which no assembly occurs in bulk solution.

Going beyond this simple thermodynamic analysis, the statistical mechanics of budding of a preassembled nucleocapsid driven by interactions with membrane-associated GPs (Fig.~\ref{fig:buddingSchematic}B) was considered in continuum models by Refs. \cite{Tzlil2004, Lerner1993}. Tzlil et al. \cite{Tzlil2004} identified a critical GP-capsid adhesion energy above which complete budding occurs, and that when complete budding occurs capsids are nearly saturated with GPs. Numerous other experimental, theoretical, and computational studies have analyzed the exit (exocytosis) or entry (endocytosis) of viruses and other nanoscale particles (reviewed in \cite{Barrow2013}).

\subsection{Modeling of assembly and budding dynamics.}
\label{sec:membraneDynamics}
The first analysis of the dynamics of assembly and budding by membrane-adsorbed capsid proteins was performed by Zhang and Nguyen \cite{Zhang2008}. They developed a continuum model description, in which membrane-associated partial-capsid intermediates are represented by hemispherical shells and the membrane deformation energy is modeled by the Helfrich free energy, Eq.\eqref{eq:Ebend}.  The model predicted, depending on subunit supersaturation, complete assembly and budding, or stalled partially assembled and partially wrapped states.

More recently, Matthews and Likos performed particle-based simulations of patchy spheres assembling into icosahedrons on a triangulated representation of a membrane (Fig.~\ref{fig:buddingSchematic}C) \cite{Matthews2012, Matthews2013a}. They found that subunit adsorption onto the membrane could drive assembly for parameters where no assembly occurred in bulk (see Eq.~\ref{eq:fcMembrane}). Interestingly, the ability of the membrane to promote assembly depended non-monotonically on the bending modulus $\kappa$, with budding suppressed by membrane bending energy at high $\kappa$  or entropic membrane fluctuations at low $\kappa$ \cite{Matthews2012}. Ruiz-Herrero and Hagan ~\cite{Ruiz-Herrero2014} considered an implicit solvent lipid bilayer membrane model and pentameric subunits that adsorb onto the membrane and assemble to form a dodecahedron (Fig.~\ref{fig:buddingSchematic}D). They found that, while adsorption onto the membrane promoted the formation of small partial capsids, the geometry of adsorbed subunits could introduce new barriers to assembly. Based on the observation that many enveloped viruses preferentially bud from lipid rafts \cite{Waheed2010, Welsch2007,Rossman2011}, assembly was also simulated on a phase separated membrane. Assembly and budding were significantly enhanced for certain domain sizes.

\section{Small molecule assembly effectors as antiviral agents}
\label{sec:antiviral}
In this section, we briefly consider small molecule `assembly effectors' which modulate assembly by perturbing capsid protein-protein interactions, altering protein conformations, or crosslinking protein-NA binding domains (for reviews see \cite{Prevelige2011,Zlotnick2011}). These molecules have been proposed as a new class of antiviral agents that interfere with capsid assembly, genome packaging, or disassembly.  This strategy has significant promise, since relatively few existing treatments target these steps of the viral life cycle.

It has been demonstrated that small molecules can stabilize the capsids of picornaviruses, inhibiting disassembly and preventing viral propagation \cite{Plevka2013, Colibus2014}. Directly interfering with assembly may also be a viable strategy; for HIV, peptides and small molecules have been developed which interfere with subunit-subunit interaction and prevent assembly in vitro, as well as trigger premature disassembly and inhibit capsid maturation \cite{Prevelige2011}. Interestingly, two classes of small molecules have been found to interfere with HBV capsid assembly by {\bf strengthening} the interaction between capsid subunits \cite{Katen2010, Katen2013}. The heteroaryldihydropyrimidine (HAP) compounds both strengthen and subtly alter the geometry of the HBV capsid protein subunit-subunit interaction, leading to malformed capsids (see section~\ref{sec:emptyCapsidProducts}). The phenylpropenamides compounds do not alter the capsid geometry, but by strengthening subunit-subunit interactions they enable the assembly in the absence of the genome (see section~\ref{sec:cargoThermo}), leading to empty capsids, which are (likely) a dead end for the HBV viral life cycle.

\section{Outlook}
\label{sec:outlook}
We have presented a summary of capsid assembly mechanisms, how they are influenced by non-protein components such as nucleic acids and lipid bilayers, and some virus-specific interactions that facilitate selective genome packaging.  Because most intermediates on assembly pathways remain challenging to characterize, complementary experimental and theoretical investigations will play important roles in further elucidating assembly mechanisms. As computer power and the sophistication of computational methods increase, models with increased resolution and complexity will become feasible. At the same time, new experimental capabilities to monitor the assembly of individual viruses will place additional constraints with which to more stringently test and refine these models.

For biomedical applications, it will be essential to build models that incorporate factors specific to host environments, such as molecular crowding \cite{Smith2014}, subcellular localization \cite{Bamunusinghe2011}, coordinated translation and assembly \cite{Dykeman2014, Annamalai2008,Kao2011}, and active processes \cite{Taylor2011,Gladnikoff2009,Balasubramaniam2011a}. Doing so will require more quantitative data from in vivo experiments and  in vitro systems that systematically incorporate host factors, in conjunction with corresponding comprehensive models. Such combined experiments and models can reveal the features that make virus assembly so robust, and identify the factors which are most sensitive to manipulation by drugs or other antiviral agents.

{\bf Acknowledgement.} This work was supported by the NIH through Award Number R01GM108021 from the National Institute Of General Medical Sciences and the NSF through award number NSF-MRSEC-0820492. We thank Chuck Knobler for a critical read of the manuscript.

\bibliographystyle{rlj}
\bibliography{all-references}

\end{document}